\documentclass[12pt,preprint]{aastex}

\usepackage{xspace}
\usepackage[]{natbib}

\newcommand{\myemail}{kenji@milkyway.gsfc.nasa.gov}
\newcommand{\OBSP}{Obs$_{\rm XMM1}$\xspace}
\newcommand{\OBSS}{Obs$_{\rm XMM2}$\xspace}
\newcommand{\OBSPS}{Obs$_{\rm XMM1\&2}$\xspace}

\newcommand{\OBSCA}{Obs$_{\rm CXO1}$\xspace}
\newcommand{\OBSCB}{Obs$_{\rm CXO2}$\xspace}
\newcommand{\OBSCAB}{Obs$_{\rm CXO1\&2}$\xspace}

\newcommand{\XB}{X$_{{\rm E}}$\xspace}
\newcommand{\XA}{X$_{\rm W}$\xspace}

\newcommand{\ASCA}{{\it ASCA}\xspace}

\newcommand{\HIPPARCOS}{{\it HIPPARCOS}\xspace}

\newcommand{\XMM}{{\it XMM-Newton}\xspace}

\newcommand{\CHANDRA}{{\it Chandra}\xspace}

\newcommand{\UNITFLUX}{{\rm ergs~cm$^{-2}$~s$^{-1}$}\xspace}

\newcommand{\UNITCPS}{{\rm cnts~s$^{-1}$}\xspace}

\newcommand{\UNITPFLUX}{{\rm cnts~cm$^{-2}$~s$^{-1}$}\xspace}
\newcommand{\UNITLUMI}{{\rm ergs~s$^{-1}$}\xspace}
\newcommand{\UNITEI}{{\rm cm$^{-3}$}\xspace}

\newcommand{\UNITNH}{{\rm cm$^{-2}$}\xspace}
\newcommand{\UNITVEL}{{\rm km~s$^{-1}$}\xspace}

\newcommand{\DEGREE}{{$^{\circ}$}\xspace}
\newcommand{\ARCMIN}{{$'$}\xspace}
\newcommand{\FARCM}{{.$'$}\xspace}
\newcommand{\ARCSEC}{{$''$}\xspace}
\newcommand{\FARCS}{.{$''$}\xspace}

\newcommand{\FSECM}{.{$^{s}$}\xspace}

\newcommand{\NH}{{\it N$_{\rm H}$}\xspace}
\newcommand{\AV}{{\it A$_{\rm V}$}\xspace}

\newcommand{\LX}{{\it L$_{\rm X}$}\xspace}

\newcommand{\KT}{{\it kT}\xspace}

\newcommand{\EM}{{\it E.M.}\xspace}

\newcommand{\PSF}{{\it psf}\xspace}

\slugcomment{Revision according to the second referee's comments}

\shorttitle{Extremely Embedded X-ray Sources in the R CrA Core}
\shortauthors{Hamaguchi et al.}

\begin{document}

\title{Discovery of Extremely Embedded X-ray Sources in the R Coronae Australis Star
Forming Core}

\author{Kenji Hamaguchi\altaffilmark{1}, Michael F. Corcoran\altaffilmark{2}, Rob Petre, Nicholas E. White}
\affil{Exploration of the Universe Division, Goddard Space Flight Center,
Greenbelt, MD 20771, USA}
\email{\myemail; corcoran@lheapop.gsfc.nasa.gov; robert.Petre-1@nasa.gov; nwhite@lheapop.gsfc.nasa.gov}

\and 

\author{Beate Stelzer}
\affil{INAF, Osservatorio Astronomico di Palermo, Piazza del Parlamento 1, I-90134 Palermo, Italy}
\email{stelzer@astropa.unipa.it}

\and 

\author{Ko Nedachi, Naoto Kobayashi}
\affil{Institute of Astronomy, University of Tokyo, 2-21-1 Osawa, Mitaka, Tokyo 181-0015, Japan}
\email{nedachi@subaru.naoj.org; naoto@ioa.s.u-tokyo.ac.jp}

\and

\author{Alan T. Tokunaga}
\affil{Institute for Astronomy, University of Hawaii, 2680 Woodlawn Drive, Honolulu, HI 96822, USA}
\email{tokunaga@ifa.hawaii.edu}

\altaffiltext{1}{National Research Council, 500 Fifth Street, NW, Washington, D.C. 20001, USA}
\altaffiltext{2}{Universities Space Research Association, 7501 Forbes 
Blvd, Ste 206, Seabrook, MD 20706, USA}

\begin{abstract}
With the  \XMM and \CHANDRA observatories,
we detected two extremely embedded X-ray sources in the R Corona Australis (R~CrA)
star forming core, near IRS~7.
These sources, designated as \XB and \XA, have
X-ray absorption columns of 
$\sim$3$\times$10$^{23}$ \UNITNH equivalent to \AV $\sim$180$^{m}$. They 
are associated with the VLA centimeter radio sources 10E and 10W, respectively.
\XA is the counterpart of the near-infrared source IRS~7, whereas
\XB has no $K$-band counterpart above 19.4$^{m}$.
This indicates that \XB is younger than typical Class~I protostars, probably a Class~0 protostar 
or in an intermediate phase between Class~0 and Class~I.
The X-ray luminosity of \XB varied between 29$<\log$ \LX $<$31.2  \UNITLUMI on timescales
of 3--30 months.  \XB also showed a monotonic increase in X-ray brightness by a factor of two in 30~ksec during an \XMM observation.
The \XMM spectra  indicate emission from a hot plasma with 
\KT $\sim$3--4~keV and also show fluorescent emission from  cold iron.
Though the X-ray spectrum from \XB is similar to flare spectra from
Class~I protostars in luminosity and temperature, 
the light curve does not resemble
the lightcurves of magnetically generated X-ray flares because the variability timescale of \XB is too long and because
variations in X-ray count rate were not accompanied by
variations in spectral hardness.
The short-term variation of \XB may be caused by the partial blocking 
of the X-ray plasma, while the month-long flux enhancement
may be driven by mass accretion.
\end{abstract}
\keywords{stars: activity --- stars: magnetic fields --- stars: pre--main-sequence --- stars: rotation --- X-rays: stars}

\section{Introduction}
Low-mass protostars are divided into two classes, Class 0 and Class I, according to their infrared (IR) and radio spectral energy distribution (SED).
This classification generally traces their evolutionary status.
Class 0 objects are thought to be young ($t\sim$10$^{4}$~yr) protostars which mainly emit in the far-IR 
and submillimeter wavelengths with blackbody temperatures of $<$30~$K$ \citep[][]{Andre1993}.
They are believed to be accreting mass dynamically from their huge circumstellar envelopes.
Class I objects are believed to be older protostars ($t \sim$10$^{5}$~yr) at the end of the mass-accretion phase and emit in the near-IR at temperatures of 
$T \sim$3000$-$5000~$K$.

Protostellar cores are generally hidden inside enormous gas envelopes.
Hard X-rays can penetrate the thick molecular clouds, and hard X-ray observations
have revealed high energy activity 
associated with Class~I objects
\citep{Koyama1996,Grosso1997,Imanishi2001}.
The observed X-ray emission exhibits occasional rapid outbursts reminiscent of solar flares
though current star formation theories do not predict solar-type magnetic dynamos
in very young stars.
\citet{Montmerle2000} proposed an alternative dynamo 
mechanism, in which fossil magnetic fields link 
the protostellar core with its circumstellar disk reconnect.

Hard X-ray emission from other embedded sources 
was reported in the OMC-2/3 cloud \citep{Tsuboi2001}.
The two detected sources show Class~0 characteristics:
huge absorption (\NH $\sim 1-3 \times 10^{23}$\UNITNH), no near-IR counterparts, and 
associations with millimeter radio clumps.
However, follow-up radio and near-IR observations by \citet{Tsujimoto2004} 
did not unambiguously classify them as Class~0 protostars.
In particular, one of these sources correlates with a centimeter radio 
source and a jet feature in the H$_{2}$ band, which indicates excitation by a jet 
from a nearby Class~I protostar.
Though \citet{Skinner2003} and \citet{Rho2004} have reported 
X-ray emission from other millimeter radio clumps in NGC~2024 and the Trifid nebula,
the photon statistics in those observations were  
too limited ($\lesssim$50 photons per source) to identify their nature conclusively.
To date no X-ray source has been clearly identified with a bona-fide Class~0 object. 

The R Corona Australis (R~CrA) cloud is a nearby 
star forming region \citep[$d \sim$170 pc,][]{Knude1998}.
Among many young stellar objects in the cloud,
those in the IRS~7 region have attracted particular interest as a site of ongoing star formation.
The region contains double peaked, strong centimeter emission; an eastern peak is
designated as 10E or IRS~7B, and a western peak
is
designated 10W or IRS~7A \citep[][]{Brown1987,Feigelson1998,Harju2001}.
It also contains two submillimeter peaks \citep{Ancker1999b}, multiple
millimeter continuum peaks \citep{Henning1994,Saraceno1996,Chini2003,Choi2004}, and a
signature of strong bipolar outflows \citep{Harju1993,Anderson1997},
but only one near-IR source,
IRS~7 \citep[][W97]{Wilking1997}.
These characteristics make the IRS~7 region a promising host of Class~0 sources.
\citet{Koyama1996} detected hard X-ray emission and an intense flare from the IRS~7 
region, which suggested the presence of a magnetically active protostar. 

In this paper, we report 
the X-ray detection of two extremely embedded sources in the IRS~7 star
forming core during \XMM observations in 2003.
These results, combined with the analysis of two \CHANDRA observations carried out
in 2000 and 2003, and
follow-up near-IR observations with the University of Hawaii 88-inch Telescope (UH88),
help to determine the nature of the detected X-ray sources.
 
\section{X-ray Observations and Absolute Position Correction}

We analyzed X-ray data obtained with the \XMM observatory 
on \dataset[14690101]{2003 March 28} and 
\dataset[14690201]{2003 March 29} (hereafter \OBSP and \OBSS, see Table \ref{tbl:obslogs} for details).
The \XMM satellite \citep{Jansen2001} is composed of three nested Wolter I type X-ray telescopes \citep{Aschenbach2000} with European Photon Imaging Camera (EPIC) CCD detectors 
in their focal planes \citep[][]{Struder2001, Turner2001}. 
The observations were obtained with the EPIC pn and EPIC MOS1 and EPIC MOS2 detectors,
in full frame mode with the medium filter.
We pointed at similar sky positions in \OBSP and \OBSS;
the IRS~7 star forming core was at 6\ARCMIN off-axis,
where the 90\% photon radius is $\sim$1\ARCMIN.
We analyzed the X-ray data using
the software package SAS ver. 5.4.1 and HEAsoft ver. 5.2.
We first processed the data using the SAS scripts ``emchain" and ``epchain", and
screened out high background periods 
using standard criteria,
excluding events when 
the entire chip count rate of ``pattern = 0" events
above 10 keV 
was $>$0.35 \UNITCPS for MOS and $>$1 \UNITCPS for pn, and excluding  
events close to hot pixels or outside the field of view.
Finally, we selected events with pattern $\leqq$4 for spectral analysis of the EPIC pn and
with pattern $\leqq$12 for the EPIC MOS and the other analyses of the EPIC pn.

To help source identification and investigate long-term flux variations,
we analyzed \dataset[200017,200194]{two archival \CHANDRA observations}. 
\CHANDRA has a single high-performance X-ray telescope which provides
sub-arcsecond imaging \citep{Weisskopf2002}.
The \CHANDRA observations were obtained using the Advanced CCD Imaging Spectrometer detector using the Imaging array (ACIS-I)
on 2000 October 7 and 2003 June 26
(hereafter \OBSCA and \OBSCB, see Table \ref{tbl:obslogs}).
\citet{Garmire2002} presented the first results of \OBSCA.
The IRS~7 star forming region was at 2\ARCMIN and 0\FARCM2 off-axis 
in \OBSCA and \OBSCB, respectively, where the 90\% photon radius is $\sim$1\ARCSEC.
We analyzed the \CHANDRA data using 
the software packages CIAO ver. 2.3 and HEAsoft ver. 5.2.
For \OBSCA,
we reprocessed the level 1 event data with ``acis\_process\_events" to compensate for
degradation of the CCD spectral resolution by charge transfer inefficiency
\citep[CTI,][]{Townsley2000ia}. For \OBSCB, we used 
pipeline-processed level 2 event data 
which was already corrected for CTI degradation.
Finally, we selected events from
both observations with the standard grades 0, 2, 3, 4, and 6 
in the \ASCA event grade system for further analysis.

For the correction of the absolute coordinates,
we measured positions of X-ray bright sources on the 
combined EPIC pn plus MOS images 
between 0.3$-$10 keV and positions of sources in 
the ACIS-I image between 0.5$-$9 keV using available
source detection packages (SAS: {\it edetect\_chain}, CIAO: {\it wavdetect},
see also Table~\ref{tbl:obslogs}).
We correlated these positions with near-IR source positions in the
2MASS All-Sky Point Source Catalog (PSC) obtained from the NASA/IPAC Infrared Science Archive (GATOR)\footnote{http://irsa.ipac.caltech.edu/applications/Gator/}, whose
astrometric accuracy is good to about 0\FARCS1\footnote{Sect. 2.2 of Explanatory Supplement to the 2MASS All-Sky Data Release: http://www.ipac.caltech.edu/2mass/releases/allsky/doc/}.
Each X-ray source has only one 2MASS counterpart within 4\ARCSEC, so that
the source identification is very reliable.
After adjusting the X-ray positions to the 2MASS 
positions, the deviation of the \CHANDRA and \XMM coordinates from the 2MASS 
coordinates is $\lesssim$ 0\FARCS6.
 
\section{X-ray Image}
\label{sec:img}
\begin{figure}
\plotone{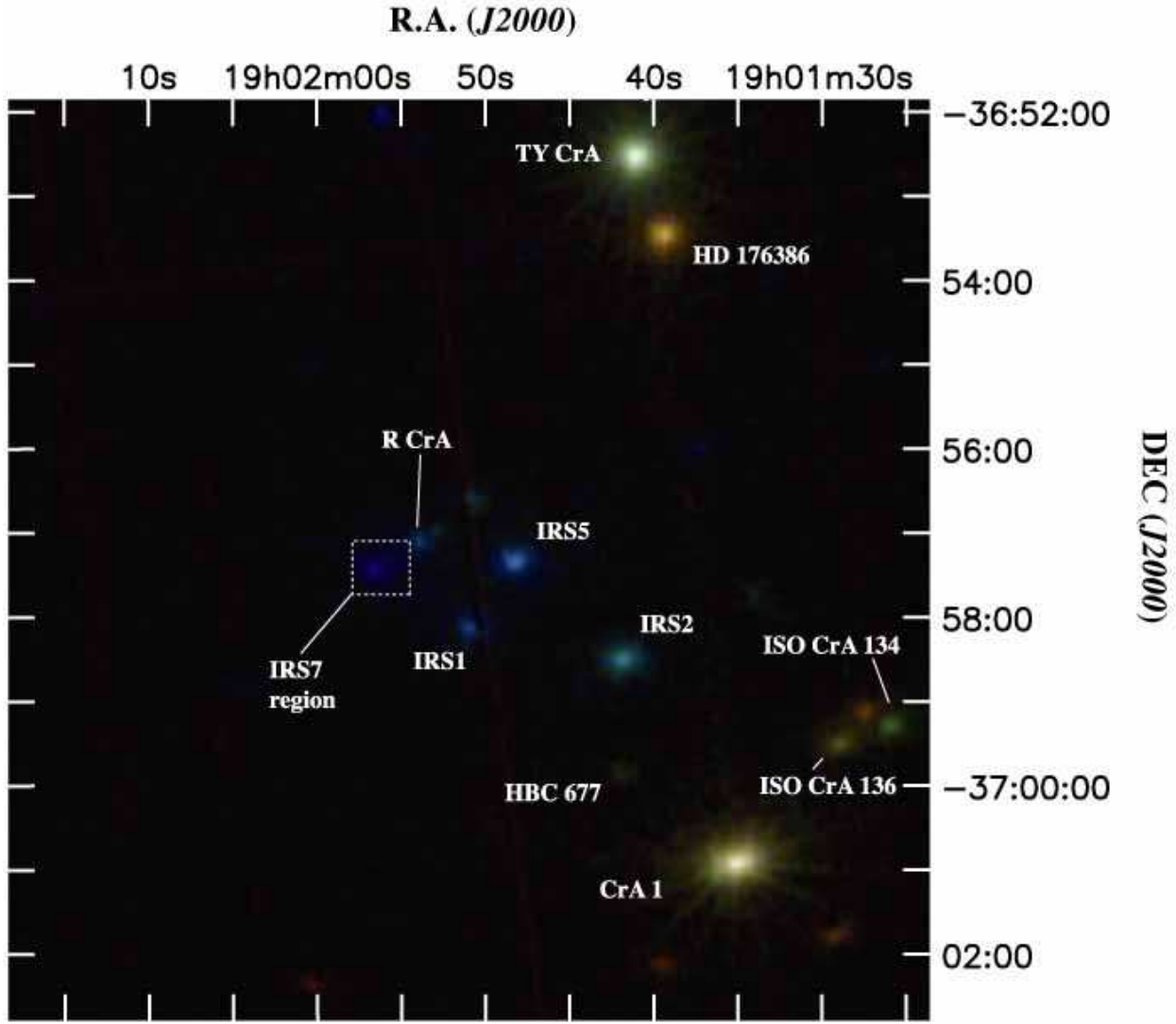}
\caption{\XMM ``true-color" image of the R~CrA star forming region (\OBSP + \OBSS). \label{imagea}
The image is color coded to represent hard band (3$-$9 keV) to blue, medium band (1$-$3 keV) to green, and soft band (0.2$-$1 keV) to red.
The dotted rectangle shows the IRS~7 region (field of view of Figure~\ref{imageb} except the bottom right panel).
Class~I protostar: IRS~1, IRS~2, IRS~5, Herbig Ae/Be star: R~CrA, TY~CrA, HD~176386, Weak-lined T-Tauri star: CrA~1.
}
\end{figure}

\begin{figure}
\plotone{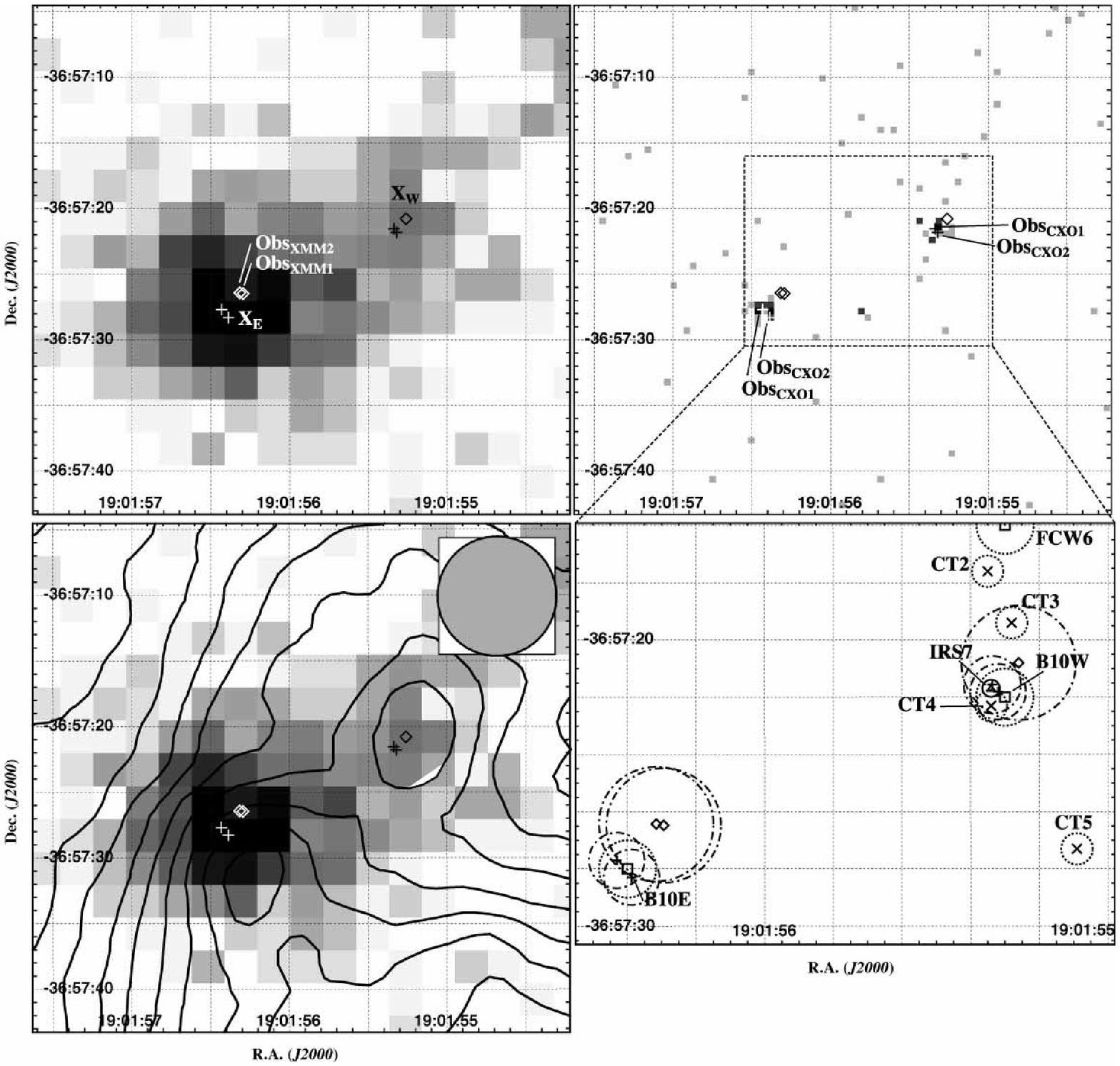}
\caption{Magnified images of the IRS~7 region. \label{imageb}
The top left panel: the 3$-$9 keV EPIC pn + MOS image of \XMM combining \OBSP and \OBSS,
the top right panel: the 0.5$-$9 keV image of \CHANDRA in \OBSCA,
the bottom left panel: the \XMM image overlaying flux contours in sub-millimeter at 450 $\mu$m 
taken by \citet{Ancker1999b},
the bottom right panel: source positions with their error circles including sources in 
near-IR and radio wavelengths.
Diamonds show \XMM sources, crosses \CHANDRA sources,
squares centimeter radio sources (B{\it number}: \citealt{Brown1987}, FCW{\it number}: \citealt{Feigelson1998}, 
Source positions are referred to \citealt{Feigelson1998}), x millimeter radio sources 
\citep[CT{\it number}:][]{Choi2004}, a triangle near-IR source (IRS~7).
The gray circle in the bottom left panel shows the beam size of the sub-millimeter telescope.
Error circles in the bottom right panel are drown in dot-bar line for X-ray sources, dotted line 
for radio sources, and solid line for the near-IR source IRS~7.
}
\end{figure}

Figure~\ref{imagea} shows a ``true-color'' X-ray image of the R~CrA star forming region, which 
combines EPIC pn and MOS(1+2) data taken in both \OBSP and \OBSS.
Red, green, and blue colors in the image represent soft (0.2$-$1 keV), medium (1$-$3 keV), and 
hard (3$-$9 keV) bands, respectively.
The IRS~7 region is colored in dark blue, which indicates the presence of 
hard X-ray sources, possibly suffering strong low-energy absorption.
The top left panel of Figure~\ref{imageb} 
shows a detailed view of the IRS~7 region of the hard band image.
A strong X-ray source, labelled \XB, was detected with {\it edetect\_chain}
in both \OBSP and \OBSS (Table \ref{tbl:detsources}).
Though {\it edetect\_chain} failed to detect a small peak on the 
tail of the \PSF to the north-west of \XB, our subsequent
analysis using the \OBSP data confirmed it as a weak source at
$\sim$14$\sigma$ significance which we labelled \XA.
The positional accuracy of both sources is hard to estimate because 
the sources are unresolved. In this paper,
we tentatively put the errors at $\sim$2\ARCSEC, the half pixel size of the EPIC pn camera.
If we rely on the result of {\it edetect\_chain},
the absolute position accuracy of \XB is 1\FARCS7 for \OBSP and 1\FARCS2 for \OBSS
at 90\% confidence.

\CHANDRA detected two weak X-ray sources in the IRS~7 region at above 4$\sigma$ significance
in both \OBSCA and \OBSCB (top right panel of Figure~\ref{imageb} for \OBSCA,
Table \ref{tbl:detsources}).
Though each source has less than 20 photons,
the absolute source positions are determined within 1\ARCSEC thanks to the
excellent spatial resolution of \CHANDRA.
These detected sources have corresponding \XMM sources though, if we take 
the error circles derived using the {\it edetect\_chain} results,
the source positions of \XB between the \CHANDRA and \XMM
observations have a significant  discrepancy of $\sim$2\ARCSEC.
Hereafter, we refer to the \CHANDRA detected positions for both sources.

The bottom panels of Figure~\ref{imageb} compare the X-ray source positions with
sources detected at other wavelengths.
\XB is associated with the VLA centimeter source 10E \citep{Brown1987}
and the sub-millimeter peak vdA~5 \citep{Ancker1999b}.
\cite{Choi2004} also suggested a marginal millimeter source at the position of \XB.
\XA is associated with centimeter 
\citep[10W in][]{Brown1987}, millimeter \citep[source 4 in][]{Choi2004}, and
sub-millimeter sources,
and the near-IR source, IRS~7 (see also section \ref{sec:irfollowup}).

\section{X-ray Light Curves and Spectra}
\label{sec:lcspec}

\subsection{East Source (\XB)}
\label{subsec:lcspecxmmxb}

We extracted source photons from the \XMM observations of \XB 
from a 27\FARCS5 radius circle centered on the 
source position excluding the region around \XA,
with background taken from appropriate source free regions.
We obtained $\sim$2,000 net counts from \OBSP and \OBSS, 
enabling us to perform detailed timing and spectral analyses.
For the \CHANDRA data, we extracted events from a 1\FARCS8 radius circle 
centered on the source position.
The background level was negligibly low.
Only 10$-$20 counts were extracted from the \CHANDRA observation,
which precludes detailed spectral analysis.

\begin{figure}
\plotone{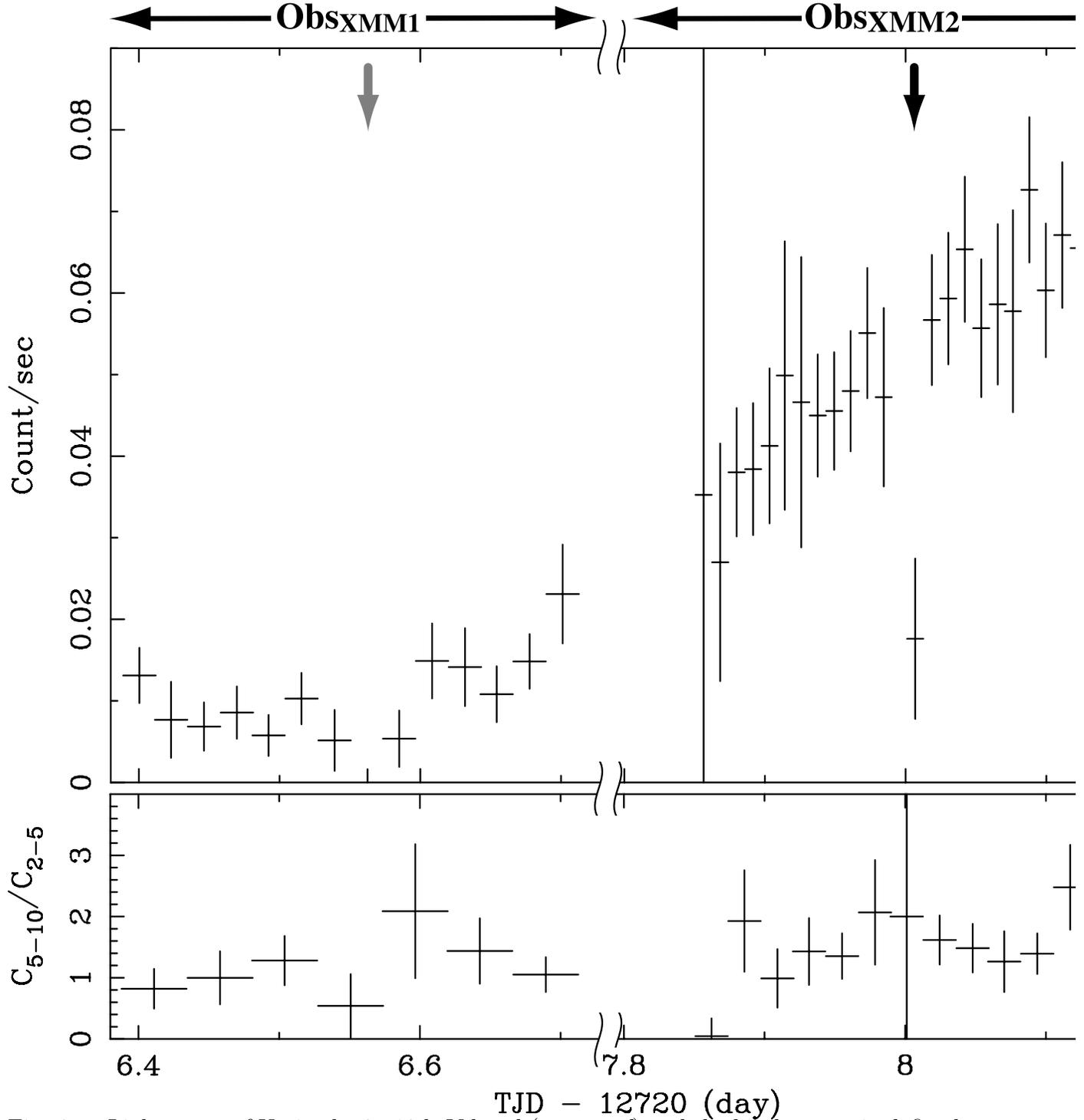}
\caption{Light curve of \XB in the 2$-$10 keV band ({\it top panel})
and the hardness ratio defined as ``count rates in the 5$-$10 keV band / count rates in the 2$-$5 keV
band"
({\it bottom panel}). \label{cur:xmmxb}
Points with error bars in the top and bottom panels are made from summed data sets of 
EPIC pn plus MOS (1+2).
The horizontal axis is truncated Julian day (TJD) $-$ 12720.
The arrows show timings of the dip feature.
Bins in the top panel have 2 ksec for \OBSP and 1 ksec for \OBSS, and bins in the 
bottom panel have 4 ksec for \OBSP and 2 ksec for \OBSS.}
\end{figure}

The top panel of Figure~\ref{cur:xmmxb} shows the background subtracted 
EPIC pn+MOS light curve of \XB in the 2$-$10 keV band.
The first half of the light curve, corresponding to \OBSP, is mostly flat with some indications of
a slight increase at the end.
A constant provides an acceptable 
$\chi^{2}$ fit at above 90\% confidence (Table \ref{tbl:lcxmmxb}).
The source was about four times brighter than the average count rate of \OBSP 
at the beginning of the second half, corresponding to \OBSS,
and the count rate gradually increased by a factor of two.
This part of the light curve can be acceptably fit at greater than 90\% confidence level 
 by a linear increase
with a slope of $\sim$9.3 $\times$10$^{-2}$ \UNITCPS day$^{-1}$(Table \ref{tbl:lcxmmxb}).
During both observations,
the hardness ratio defined as count rates in the 5$-$10 keV band over those in the 2$-$5 keV band
remained unchanged, except for a minimum at 7.86 days. 
This means that time variation in count rate was not accompanied by any
significant change of the spectral shape.
On the other hand,
the light curves show marginal dips on timescales of $<$1 ksec 
near 8.02 and 8.17 days.
These dips are seen in both the pn and MOS light curves of \XB
and other nearby sources, such as R~CrA, IRS~5, and CrA~1,
have no dips at those times.
These facts do not support an instrumental origin.
We note that these dips might suggest a tentative period of $\sim$13.9 ksec, with the
combination of minimal dips at 6.56 and 7.85 days (the dip at 7.85~day was only covered with 
the MOS data and does not appear in the Figure~\ref{cur:xmmxb}).

\begin{figure}
\plotone{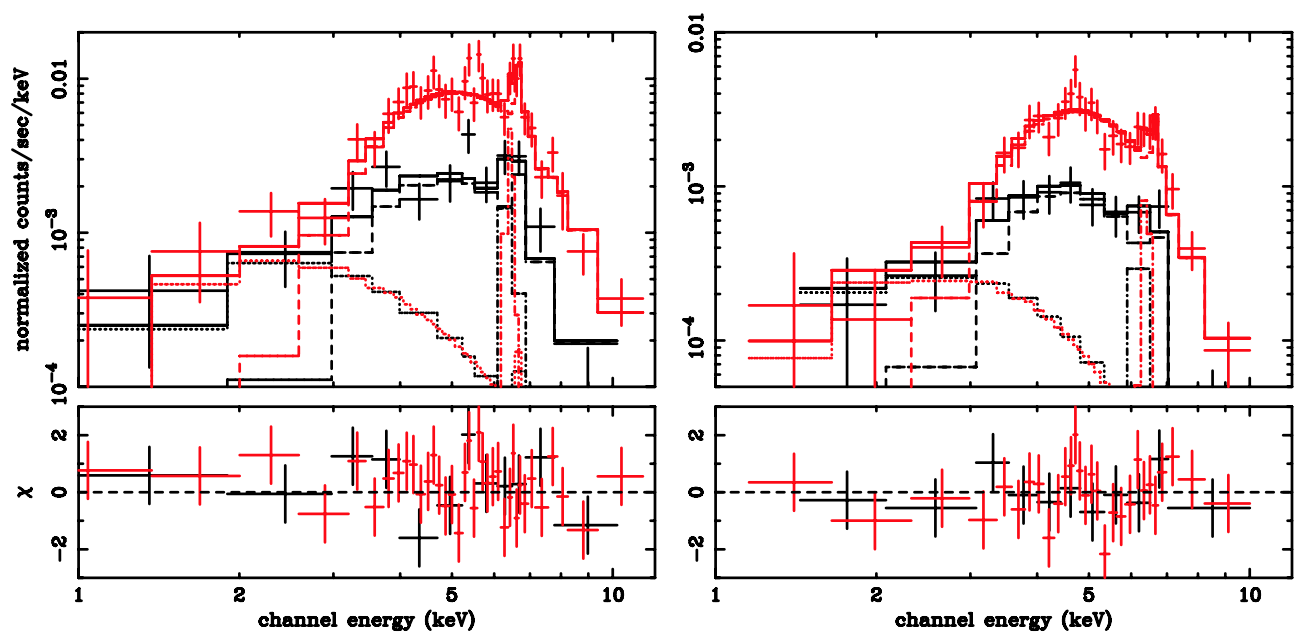}
\caption{EPIC pn ({\it left}) and MOS (1+2) ({\it right}) spectra of \XB in \OBSP ({\it black}) and \OBSS ({\it red}). \label{spec:xmmxb}
The solid lines show the best-fit model of the simultaneous fitting with EPIC pn and MOS.
Dotted lines show the soft component, barred line the hard component, and
dot-bar lines the Gaussian component for the line at 6.4~keV.
Each bottom panel shows the residuals of the $\chi^{2}$ fit.}
\end{figure}

The EPIC spectra of \XB in \OBSP and \OBSS (Figure~\ref{spec:xmmxb})  show several similarities:
significant emission up to $\sim$10 keV; strong absorption below 2$-$3 keV;
a broad line feature between 6$-$7 keV; marginal lines between 5$-$6 keV
in the EPIC pn spectra (which may be of
instrumental or cosmic origin\footnote{Sect. 3.3.7.2 in \XMM Users Handbook,\\
http://xmm.vilspa.esa.es/external/xmm\_user\_support/documentation/uhb/index..html}).
To investigate the 6$-$7~keV line feature, we fit the EPIC pn and MOS (1+2) 
spectra simultaneously with an absorbed power-law model with a Gaussian component. 
An acceptable fit above 90\% confidence has a photon index of 3.0 (2.5--3.4), 
a Gaussian line centroid of 6.60 (6.53--6.67) keV, and a Gaussian width of 0.15 (0.079--0.28) keV,
where the numbers in parentheses denote the 90\% confidence range.
The derived Gaussian width, equivalent to $\Delta v \sim$7,000 \UNITVEL if 
produced by Doppler broadening, is unreasonably large for
a stellar plasma (see also discussion \ref{subsec:naturexmmxb}).
We therefore interpret the broad feature as a blend of iron lines from a hot plasma 
at 6.7~keV and a fluorescent iron line at 6.4 keV
though the profile needs to be confirmed with deeper observations.
We then 
fit the \XMM spectra with XSPEC by an absorbed 1-temperature (1T) optically thin, thermal plasma model 
(wabs: \citealt{Morrison1983}; MeKaL: \citealt{Mewe1995})
with a Gaussian component with line center fixed at 6.4 keV.
For either of the two observations this model yields acceptable fits at the
90\,\% confidence level (Model A and B in Table \ref{tbl:spec}). 
\NH differs significantly between \OBSP and \OBSS,
possibly due to the simplistic spectral model we assumed.
Indeed, the soft emission below 3~keV was unchanged between \OBSP and \OBSS, suggesting
an additional component along with the hard emission.
In the single temperature fit to \OBSP, the soft emission is included as a part of the absorption slope
whereas the fit to \OBSS determines \NH from the 3$-$5 keV slope and 
does not reflect the soft emission.
We therefore refit the spectra of \OBSP and \OBSS simultaneously with
an absorbed 2T model 
--- 1T for the variable hard component and 1T for the constant soft component --- 
with a Gaussian line at 6.4~keV.
In this model, we tied the \NH of the hard components in \OBSP and \OBSS and tied
the elemental abundances of all components.
We allowed \NH of the soft and hard components to be fit independently
because a model fit with a common \NH gives large \NH $\sim$2.4$\times$10$^{23}$ \UNITNH and
hence an unrealistically large intrinsic log \LX $\sim$ 35 \UNITLUMI for the soft component.
The model, again, successfully reproduced the spectra
above the 90\% confidence level (Model C in Table \ref{tbl:spec}).
The derived physical parameters of the hard component are at the higher end among
those of Class~I protostars \citep[e.g., see][for comparison]{Imanishi2001}: large \NH $\sim$2.8$\times$10$^{23}$
\UNITNH, equivalent to \AV $\sim$180$^{m}$ 
(using the \NH$-$\AV relation by \citet{Imanishi2001});
plasma temperature of 3--4~keV;
log \LX $\sim$30.8 \UNITLUMI in \OBSP, which further
increased to $\sim$31.2 \UNITLUMI in \OBSS; 
and a fluorescent iron line
equivalent width (EW) 
of $\sim$810 (240$-$1400) eV in \OBSP and $\sim$250 (100$-$400) eV in \OBSS.
Meanwhile, the metal abundance is $\sim$0.2 (0.1$-$0.3)~solar, 
which is typical of low-mass young stars \citep[e.g.][]{Favata2003}.

Since we have very few counts from the \CHANDRA observations,
we simply calculated a softness ratio, defined as S / (H + S), where S and H are
photon counts in the 0.5--3~keV and 3--9~keV bands, respectively.
The ratios are 0.15$\pm$0.08 (S = 3, H = 17) in \OBSCA and 0.38$\pm$0.13 (S = 5, H = 8) in \OBSCB
where the errors show 1-sigma.
The same softness ratios evaluated for the \XMM best-fit models 
after adjusting to the \CHANDRA ACIS-I response 
are 0.11 for \OBSP and 0.048 for \OBSS, 
suggesting that  the spectra have been softer during the \CHANDRA observations.
This is consistent with the picture that the hard component further declined during the
\CHANDRA observations while the soft component was unchanged.

\subsection{West Source (\XA)}

We extracted \XMM source events of \XA from an ellipse with axes 
15\ARCSEC by 10\FARCS5 elongated toward the NNE direction, excluding a region with strong 
contamination from \XB, and selected the background from a symmetrical region near \XB.
We did not use the \OBSS data, because the \XA region suffered strong contamination from \XB.
For the \CHANDRA data, we extracted source events 
from a 1\FARCS8 radius circle.
The background level was negligibly low.

\begin{figure}
\plotone{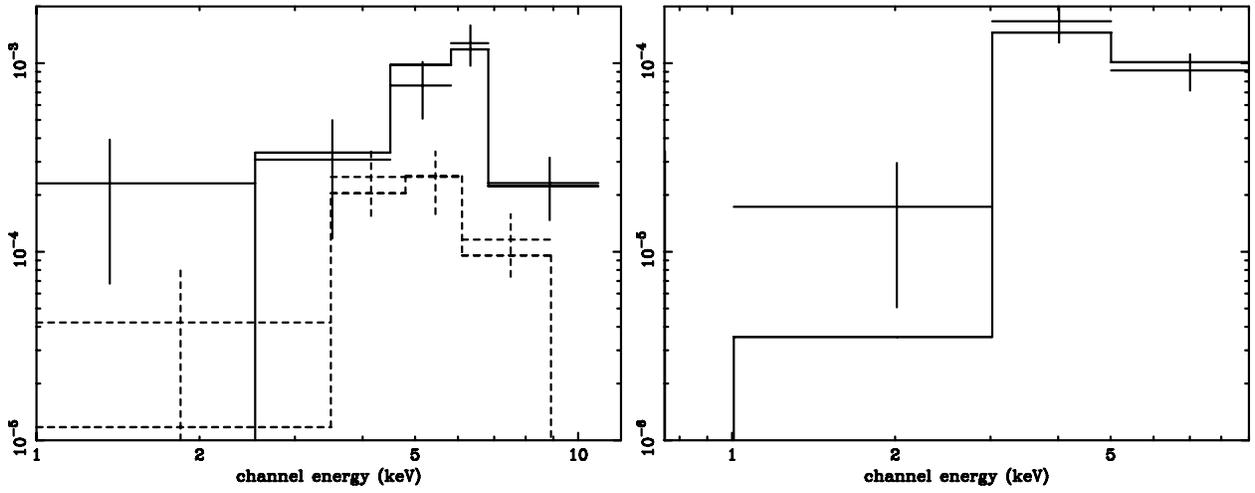}
\caption{\XA spectra in \OBSP ({\it left}, solid -- pn, dot -- MOS)
and \OBSCA plus \OBSCB ({\it right}). \label{spec:cxoxab}
Both panels show data points and their best-fit models (Model D and E in Table \ref{tbl:spec} from left).}
\end{figure}

The \XMM spectra were reproduced by an absorbed 1T model with
\NH $\sim$3.4$\times$10$^{23}$ \UNITNH,
\KT $>$1.6 keV, and log \LX $\sim$ 30.5 \UNITLUMI 
(left panel of Figure~\ref{spec:cxoxab}, Model D in Table \ref{tbl:spec}).
We added both spectra from \OBSCA and \OBSCB because their count rates are about the same.
The summed spectrum can be reproduced by the best-fit \XMM model 
just by changing its normalization (right panel of Figure~\ref{spec:cxoxab}, Model E in Table \ref{tbl:spec}).

\section{Follow-up Observations in the Near-IR}
\label{sec:irfollowup}
Infrared source catalogues in the R~CrA region currently available have several shortcomings:
shallow limiting magnitude (e.g. $K_{limit}$ $\sim$ 15$^{m}$ for 2MASS, 16.5$^{m}$ for W97); insufficient spatial resolution 
(e.g. 1\ARCSEC pix$^{-1}$ for 2MASS and 0\FARCS75 pix$^{-1}$ for W97); and
insufficient positional accuracy (e.g., $\pm$1\ARCSEC for W97).
We therefore analyzed two deep $K$-band images of the field obtained in August 1998 and August 2003
using UH88 and the near-IR imager QUIRC.
During the observations, we used
the f/10 secondary mirror, yielding a pixel scale of 0\FARCS1886 pix$^{-1}$.
The spatial resolution of our images is 0\FARCS5 and 0\FARCS8 in
FWHM for the 1998 and 2003 data respectively.
Both images were aligned with an accuracy of
0\FARCS3 by referring to the coordinates of R~CrA
in the 2MASS catalogue and a high resolution image (FWHM $\sim$0\FARCS14) obtained 
with the SUBARU telescope.
The analysis was made with IRAF\footnote{{IRAF is distributed by the National Optical Astronomy
Observatories, which are operated by the Association of Universities for Research
in Astronomy, Inc., under cooperative agreement with the National Science Foundation.}}.

\begin{figure}
\plotone{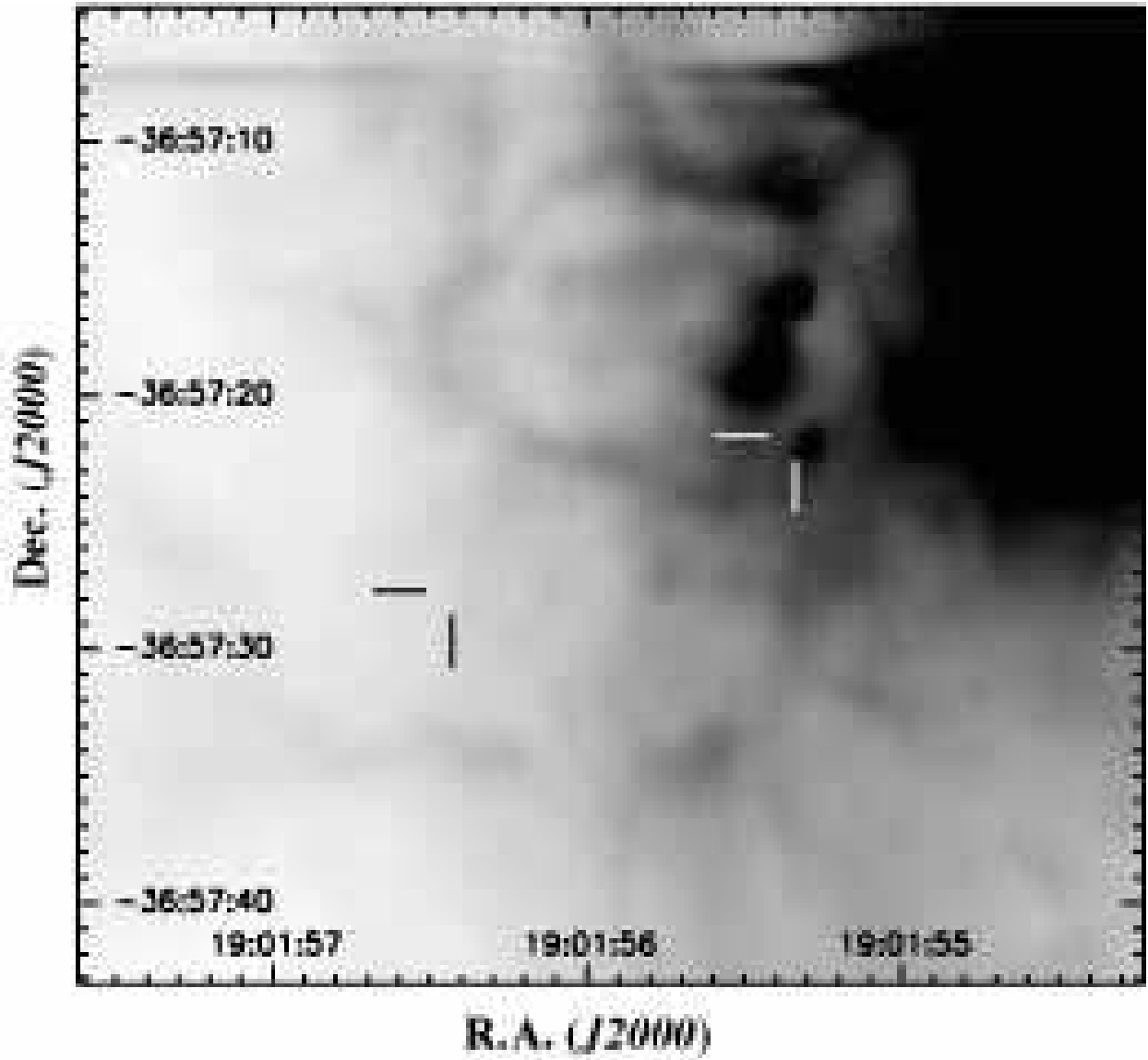}
\caption{$K$-band image of the IRS~7 region obtained in 1998.
The \XB and \XA positions measure with \CHANDRA are at the left and right crosses, respectively.\label{fig:Kbandir}
}
\end{figure}

Both images show R~CrA as a filamentary and mildly extended reflection nebulae
(Figure~\ref{fig:Kbandir} for the 1998 data), but the location of \XB 
shows no significant emission, though it does have some marginal enhancement.
The flux upper-limit was measured by subtracting the extended nebular emission, which 
we estimated with a third-order polynomial surface fit.
The 5-$\sigma$ upper-limit of \XB within the 20 $\times$ 20 pixel box 
centered at \XB was $\sim$19.4$^{m}$ for the 1998 image and $\sim$19$^{m}$ for the 2003 image.
Meanwhile, using the SUBARU image \citetext{Nedachi et al. in preparation},
we measured the absolute position of IRS~7
at ($\alpha_{2000}$, $\delta_{2000}$) = (19$^h$1$^m$55\FSECM34, $-$36\DEGREE57\ARCMIN21\FARCS69) 
with an accuracy of 0\FARCS3.
This is much more accurate than the earlier observations.
With this new position, IRS~7 falls within
the positional error circle of \XA (see the bottom right panel of Figure~\ref{imageb}).

\section{Discussion}

\subsection{The Nature and X-ray Emission Mechanism of \XB}
\label{subsec:naturexmmxb}
\citet{Harju2001} suggested that the radio counterpart of \XB, source 10E, might 
be a radio galaxy or Galactic microquasar.
However, AGNs with the observed X-ray flux above 2$-$8 $\times$10$^{-13}$ \UNITFLUX 
(2$-$10 keV) are found $\lesssim$10 degree $^{-2}$ in the sky 
\citep[][]{Ueda1998}, and
the probability to detect such an AGN in the IRS~7 star forming core ($\sim$10arcsec$^{2}$)
is extremely small ($\lesssim$10$^{-4}$).
Furthermore, AGNs do not generally show thermal iron K emission line and have rather 
flat spectral slopes ($\Gamma \leq$2) \citep[][]{Ueda1998}.
Similarly, Galactic black hole candidates
also show flat power-law 
X-ray 
spectra with $\Gamma =1.5-2.1$
though they sometimes show thermal spectra with \KT $\sim$0.5--1.5~keV
\citep{McClintock2003}.
Taking its association to the star forming core into account,
\XB is most likely a very young stellar object.

Compared to Class~I objects in the R CrA star forming region such as IRS~1, 2, and 5 
which have \NH $\sim$2$\times$10$^{22}$ \UNITNH and $K \lesssim$11$^{m}$
(from our additional analysis of the \OBSP and \OBSS data; see also W97),
\XB shows much larger X-ray absorption and much smaller near-IR luminosity.
\XB is also associated with the strong submillimeter condensation vdA~5 \citep{Ancker1999b}.
All these suggest that \XB is much younger than typical Class~I objects,
and that it is a Class~0 object or an object in an intermediate phase between Class~0 and Class~I.
We note that
extremely embedded sources in the OMC-2/3 cloud have similarly high absorption columns of 
\NH $\sim (1-3) \times 10^{23}$ \UNITNH 
\citep[][]{Tsuboi2001}.

\begin{figure}
\plotone{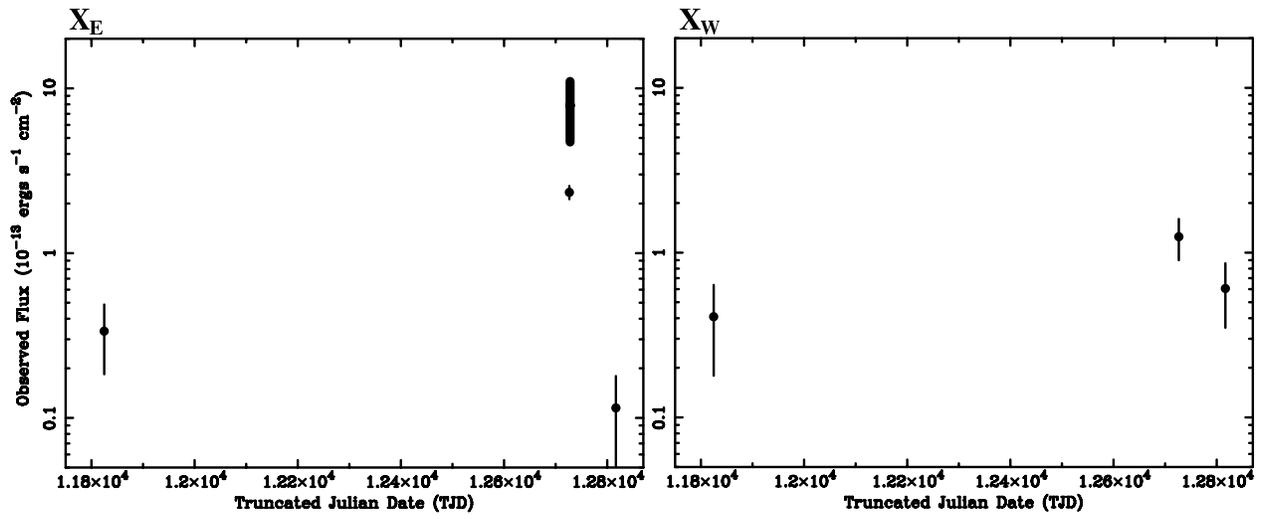}
\caption{Observed X-ray flux between 0.5--10~keV in long time scale ({\it left}: \XB, {\it right}: \XA).
The vertical narrow bars on the data points show photon statistical error at 
90\% confidence level. The thick bar shows the variable range of \XB during \OBSS.
}\label{fig:longtermvar}
\end{figure}

Between the \CHANDRA and \XMM observations,
\XB exhibited strong long-term X-ray variation by a factor of 10$-$100 on 
a timescale of a month (Figure \ref{fig:longtermvar}).
In none of the observations did we detect obvious flare activity though
\OBSS showed a marked flux increase. 
Active stars such as RS CVn and young stars in open clusters do not generally vary
in X-rays more than a factor of 2$-$3 outside flares \citep{Stern1998}.
Less active stars such as the Sun exhibit strong X-ray variations by up to a factor of 100, 
coincident with their activity cycles \cite[e.g.][]{Favata2004}, but,  unlike \XB,
the X-ray luminosity of such stars is typically less than 10$^{28}$ \UNITLUMI and the observed
activity time scale is several years.
One possibility is that the strong variability of \XB could indicate abrupt activity
produced by an enhanced mass accretion episode
similar to that recently attributed to the outburst of the star in McNeil's nebula  \citep[][]{Kastner2004}.
Indeed, the outburst increased the X-ray flux by a factor of 50, and the post-outburst 
X-ray luminosity of 10$^{31}$ \UNITLUMI 
is comparable to the luminosity of \XB during \OBSS.

The plasma temperature and X-ray luminosity of \XB exceed the typical quiescent X-ray activity 
of Class~I protostars and are comparable to temperatures and luminosities of
X-ray flares from Class~I protostars
\citep{Imanishi2001,Shibata2002}\footnote{ \citet{Imanishi2001} 
used the distance to the $\rho$ Oph cloud of 165 pc instead of 120 pc derived from more
reliable \HIPPARCOS data \citep{Knude1998} for a comparison to earlier X-ray results
of the $\rho$ Oph field. 
Their X-ray luminosity should be divided by a factor of two to compare to our result.}.
X-ray flares from Class~I protostars may be produced by reconnection in a magnetosphere which is twisted due to the core-disk differential rotation \citep{Tsuboi2000,Montmerle2000}.
Perhaps a similar mechanism explains the X-ray emission from \XB during the \XMM
observations, though magnetic reconnection would have to occur throughout the \XMM observations 
since no rapid X-ray variation was seen from \XB.

The flux increase of a factor of two in $\sim$30 ksec in \OBSS is unlike the types of variations seen in
magnetically driven X-ray flares which are characterized by rapid ($\sim$ 10 ksec) flux increases
\citep[e.g.][]{Tsuboi1998,Tsuboi2000,Stelzer2000,Imanishi2001}.
\citet{Favata2003} found a similar rise in X-ray brightness in the classical T-Tauri star XZ Tau,
with a factor of 4 increase during 50 ksec.
In this case, the brightening was accompanied by an \NH decrease and therefore \citet{Favata2003}
interpreted it as an eclipse of the emitting region by the accretion stream.
Because \XB did not show any significant hardness ratio variation
the absorber would have to be uniformly dense, optically thick gas.
Such a variation could be produced by an eclipse of the X-ray emitting region by an absorber
or emergence of the X-ray emitting region from behind
the rim of the protostellar core as a result of stellar rotation.

If the fluorescent iron line in the spectra is real,
this is unusual because fluorescent iron lines have been rarely observed 
from pre--main-sequence stars.
Even a few examples during strong flares from Class~I protostars have EW$\lesssim$ 150~eV 
\citep[][]{Imanishi2001}.
The large equivalent width of the fluorescent line from \XB ($\sim$250$-$800 eV)
again suggests that the source is extremely embedded.
When we simulate fluorescent iron line EWs, assuming solar abundance for the surrounding cold gas
\citep{Inoue1985},
an optically thick absorber should block the direct X-ray emission by
$\sim$60\% for \OBSP and $\sim$3\% for \OBSS.
This result is consistent with 
obscuration of the X-ray emission though the blocking factor
in \OBSS should be $\gtrsim$30\% to explain the observed flux increase in \OBSS.
Interestingly, the intrinsic X-ray luminosity in \OBSP should be log \LX $\sim$31.2 \UNITLUMI,
which is comparable to \LX in \OBSS.

Our thermal model fit requires a metal abundance of $\sim$0.2 (0.1--0.3) solar.
Though the derived abundance is dependent on the thermal model,
and identifying emission lines in low resolution spectroscopy may be difficult 
especially around $\sim$1~keV \citep[e.g.][]{Kastner2002b},
the metal abundance of \XB is determined mostly from the iron K line
region, which seems to show a real underabundance.
X-ray emission from T Tauri stars similarly show sub-solar abundances \citep[e.g.,][]{Favata2003}.
This might suggest that \XB has an X-ray emission mechanism 
similar to T Tauri stars.
Such abundance anomalies could be produced by 
the FIP (first ionization potential) effect \citep[e.g.][]{Gudel2001}.
Unfortunately, the spectrum of \XB does not have enough counts to detect, unambiguously,
emission lines from other elements
such as argon and calcium to test the FIP effect. 
Another possible solution is that the continuum emission includes a
non-thermal component, as proposed for $\rho$ Oph S1 \citep[][]{Hamaguchi2003}.

\begin{figure}
\plotone{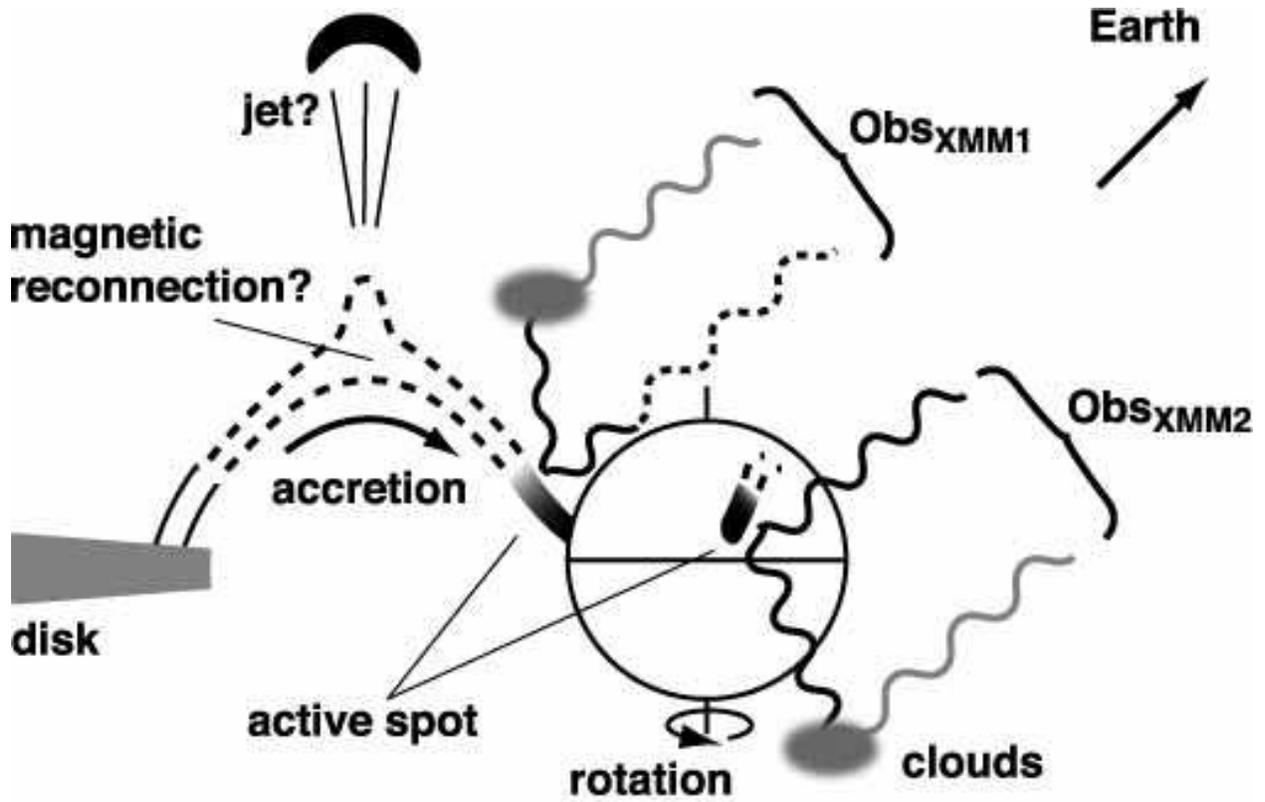}
\caption{Possible geometry of the proto-stellar core, jet, and observer \label{image:config}.
The wavy lines show X-ray emission (solid black: direct X-rays, grey: fluorescent and scattering
X-rays, dotted black: direct X-rays are partially covered).}
\end{figure}

From the above discussion,
we hypothesize that the X-ray plasma responsible for the hard component
was produced at a mass accretion spot on the protostellar core.
The X-ray emission is blocked when 
the spot would have been behind the protostellar core during \OBSP, and 
just appeared from behind the rim in \OBSS as a consequence of proto-stellar rotation
(Figure~\ref{image:config}).
To be consistent with the observed \XMM light curves, the rotational period of the
proto-stellar core would need to be $\gtrsim$2.8 days. This rotation speed 
is much slower than the break-up rotation
speculated for Class~0 protostars from rotational periods of Class~I protostars 
\citep[e.g.][]{Montmerle2000}. 

The soft component was apparently constant and had much smaller \NH 
compared to the hard component.
This may suggest that the component has no physical connection to, and exists far from, the hot component.
One possible origin is that the soft component is associated with another hidden protostar, but,
though the \NH of the soft component is typical of Class~I protostars,
the $K$-band magnitude of $\gtrsim$19$^m$ is much larger than those
of Class~I protostars in the R CrA cloud ($K \lesssim$11$^{m}$).
Another possible origin is that the X-ray plasma is heated by a collision of a
steady jet or outflow from \XB with circumstellar gas, a mechanism 
thought to be associated with X-ray emission
from HH2, L1551 IRS~5, and OMC~2/3 \citep{Pravdo2001,Favata2003,Tsujimoto2004}.
Indeed, \XB is associated with a centimeter radio source as those systems are, but
the plasma temperature and X-ray luminosity of \XB are very large compared 
to those sources, except for the source in OMC~2/3. Such a high plasma temperature requires
an energetic jet with $v_{jet} \sim$1,500 \UNITVEL.
While low-mass young stars generally have slow outflow 
velocities (a few hundred \UNITVEL),
\citet{Marti1995} measured a large proper motion in the young stellar jets HH80-81
implying velocities up to 1,400~\UNITVEL.
\XB could be another example of a source with high speed outflow.

\subsection{What is \XA?}

\XA is a counterpart to the near-IR source IRS~7 as well as the centimeter radio 
source 10W.
The column density of \XA ($\sim$2$-$6$\times$10$^{23}$ \UNITNH) is much larger than
that typical of Class~I protostars,
while the $K$-band magnitude of $\sim$12.2$^{m}$ is comparable to 
the brightness of Class~I protostars in the R CrA cloud.
Therefore, \XA might be a Class~I source seen at a large inclination angle.
On the other hand, IRS~7 had been suspected to be the counterpart of 
an X-ray flare source seen during an \ASCA observation \citep{Koyama1996}.
However, the \NH measured during the \ASCA
flare ($\sim$4$\times$10$^{22}$~\UNITNH) was
significantly smaller than that of \XA,
even considering that the \ASCA spectrum is contaminated by
emission from surrounding Class~I protostars with lower \NH.
If the \ASCA flare source is really 
\XA, the absorption column must be variable.

\section{Summary \& Conclusion}
We discovered two extremely embedded X-ray sources
at the positions of the strong VLA centimeter radio sources in the IRS~7 star forming core.
Thanks to its vicinity to the Sun ($d \sim$170 pc), the large effective area of \XMM and an
opportunity to catch an active phase,
we obtained around $\sim$2,000 photons from \XB, which is about 40 times better than
other extremely embedded X-ray sources observed so far.
The upper-limit to the $K$-band luminosity was restricted to $\sim$1/150 of the extremely
embedded protostars
in the OMC-2/3 cloud \citep[$d \sim$450 pc and completeness limit $\sim$16$^{m}$,][]{Tsujimoto2003}.
This is therefore a rare, high signal-to-noise X-ray detection of what appears to be
a Class~0 protostar or a protostar in an intermediate phase between Class~0 and Class~I.
The combination of \XMM and \CHANDRA observations 
enabled us to study the X-ray emission mechanism in detail.
The source showed a significant long-term X-ray variation by a factor of 10--100.
The X-ray properties -- \KT $\sim$3$-$4 keV and log \LX $\sim$31 \UNITLUMI --
are comparable to magnetically active X-ray flaring sources 
though the light curve of \XB does not suggest flaring activity.
These results may suggest that the X-ray activity was enhanced by mass accretion. 
The X-ray flux during \OBSS increased monotonically by a factor of two in 30~ksec,
and the X-ray spectra showed a strong iron fluorescent line with EW of 250--800~eV.
These phenomena may be related to the
partial blocking of the X-ray plasma
or perhaps 
they are 
indicative of rotation of a hot spot on the proto-stellar core.

The evolutionary status of \XB should be constrained by constructing the IR and radio SED.
This requires high spatial resolution and high sensitivity.
Spitzer, ALMA, and 8~m class ground based telescopes are well-suited
to examine this mysterious X-ray protostar.
On the other hand, long exposure X-ray observations
would confirm the fluorescent line profile and test the suggested periodicity in the X-ray brightness.
These studies would help to understand the physical nature of protostars at the earliest phase.

\acknowledgments

We are grateful to M. Choi, T. Hunter, T. Yokoyama, K. Tatematsu, M. Tsujimoto, H. Murakami,
K. Koyama, S. Yashiro, Y. Tsuboi, and the anonymous referee for useful comments.
This work is performed while K.H. held awards by
National Research Council Research Associateship Award at NASA/GSFC,
and is supported by \XMM US grant.

Facilities: \facility{XMM-Newton(EPIC)}, \facility{CXO(ACIS-I)}. \facility{UH88(QUIRC)}, 
\facility{SUBARU(IRCS+AO)}

\bibliographystyle{apj}
\bibliography{inst,sci_AI,sci_JZ,scibook}

\begin{thebibliography}{47}
\expandafter\ifx\csname natexlab\endcsname\relax\def\natexlab#1{#1}\fi

\bibitem[{{Anderson} {et~al.}(1997){Anderson}, {Harju}, {Knee}, \&
  {Haikala}}]{Anderson1997}
{Anderson}, I.~M., {Harju}, J., {Knee}, L.~B.~G., \& {Haikala}, L.~K. 1997,
  \aap, 321, 575

\bibitem[{{Andr\'{e}} {et~al.}(1993){Andr\'{e}}, {Ward-Thompson}, \&
  {Barsony}}]{Andre1993}
{Andr\'{e}}, P., {Ward-Thompson}, D., \& {Barsony}, M. 1993, \apj, 406, 122

\bibitem[{{Aschenbach} {et~al.}(2000){Aschenbach}, {Briel}, {Haberl},
  {Br\"{a}uninger}, {Burkert}, {Oppitz}, {Gondoin}, \& {Lumb}}]{Aschenbach2000}
{Aschenbach}, B., {Briel}, U.~G., {Haberl}, F., {Br\"{a}uninger}, H.~W.,
  {Burkert}, W., {Oppitz}, A., {Gondoin}, P., \& {Lumb}, D.~H. 2000, in SPIE,
  Vol. 4012, X-Ray Optics, Instruments, and Missions III, ed. {Joachim E.
  Tr\"{u}mper, Bernd Aschenbach}, p. 731--739

\bibitem[{{Brown}(1987)}]{Brown1987}
{Brown}, A. 1987, \apjl, 322, L31

\bibitem[{{Chini} {et~al.}(2003){Chini}, {K{\" a}mpgen}, {Reipurth},
  {Albrecht}, {Kreysa}, {Lemke}, {Nielbock}, {Reichertz}, {Sievers}, \&
  {Zylka}}]{Chini2003}
{Chini}, R., et al. 2003, \aap, 409, 235

\bibitem[{{Choi} \& {Tatematsu}(2004)}]{Choi2004}
{Choi}, M., \& {Tatematsu}, K. 2004, \apjl, 600, L55

\bibitem[{{Favata} {et~al.}(2003){Favata}, {Giardino}, {Micela}, {Sciortino},
  \& {Damiani}}]{Favata2003}
{Favata}, F., {Giardino}, G., {Micela}, G., {Sciortino}, S., \& {Damiani}, F.
  2003, \aap, 403, 187

\bibitem[{{Favata} {et~al.}(2004){Favata}, {Micela}, {Baliunas}, {Schmitt},
  {G{\" u}del}, {Harnden}, {}, {Sciortino}, \& {Stern}}]{Favata2004}
{Favata}, F., {Micela}, G., {Baliunas}, S.~L., {Schmitt}, J.~H.~M.~M., {G{\"
  u}del}, M., {Harnden}, {}, F.~R., {Sciortino}, S., \& {Stern}, R.~A. 2004,
  \aap, 418, L13

\bibitem[{{Feigelson} {et~al.}(1998){Feigelson}, {Carkner}, \&
  {Wilking}}]{Feigelson1998}
{Feigelson}, E.~D., {Carkner}, L., \& {Wilking}, B.~A. 1998, \apjl, 494, L215

\bibitem[{{Garmire} \& {Garmire}(2002)}]{Garmire2002}
{Garmire}, G., \& {Garmire}, A. 2002, APS Meeting Abstracts, 17064

\bibitem[{{Grosso} {et~al.}(1997){Grosso}, {Montmerle}, {Feigelson},
  {Andr\'{e}}, {Casanova}, \& {Gregorio-Hetem}}]{Grosso1997}
{Grosso}, N., {Montmerle}, T., {Feigelson}, E.~D., {Andr\'{e}}, P., {Casanova},
  S., \& {Gregorio-Hetem}, J. 1997, \nat, 387, 56

\bibitem[{{G\"{u}del} {et~al.}(2001){G\"{u}del}, {Audard}, {Smith}, {Sres},
  {Escoda}, {Wehrli}, {Guinan}, {Ribas}, {Beasley}, {Mewe}, {Raassen}, {Behar},
  \& {Magee}}]{Gudel2001}
  {G\"{u}del}, M., et al. 2001, in Proceedings of the
  12th Cambridge Workshop of Cool Stars, Stellar Systems, and the Sun: The
  Future of Cool-Star Astrophysics, ed. A.~{Brown}, T.~R. {Ayres}, \& G.~M.
  {Harper}, Boulder: Univ. of Colorado, in press

\bibitem[{{Hamaguchi} {et~al.}(2003){Hamaguchi}, {Corcoran}, \&
  {Imanishi}}]{Hamaguchi2003}
{Hamaguchi}, K., {Corcoran}, M.~F., \& {Imanishi}, K. 2003, \pasj, 55, 981

\bibitem[{{Harju} {et~al.}(1993){Harju}, {Haikala}, {Mattila}, {Mauersberger},
  {Booth}, \& {Nordh}}]{Harju1993}
{Harju}, J., {Haikala}, L.~K., {Mattila}, K., {Mauersberger}, R., {Booth},
  R.~S., \& {Nordh}, H.~L. 1993, \aap, 278, 569

\bibitem[{{Harju} {et~al.}(2001){Harju}, {Higdon}, {Lehtinen}, \&
  {Juvela}}]{Harju2001}
{Harju}, J., {Higdon}, J.~L., {Lehtinen}, K., \& {Juvela}, M. 2001, in ASP
  Conf. Ser. 235: Science with the Atacama Large Millimeter Array, ed.
  A.~{Wootten}, p. 125--129

\bibitem[{{Henning} {et~al.}(1994){Henning}, {Launhardt}, {Steinacker}, \&
  {Thamm}}]{Henning1994}
{Henning}, T., {Launhardt}, R., {Steinacker}, J., \& {Thamm}, E. 1994, \aap,
  291, 546

\bibitem[{{Imanishi} {et~al.}(2001){Imanishi}, {Koyama}, \&
  {Tsuboi}}]{Imanishi2001}
{Imanishi}, K., {Koyama}, K., \& {Tsuboi}, Y. 2001, \apj, 557, 747

\bibitem[{{Inoue}(1985)}]{Inoue1985}
{Inoue}, H. 1985, Space Science Reviews, 40, 317

\bibitem[{{Jansen} {et~al.}(2001){Jansen}, {Lumb}, {Altieri}, {Clavel}, {Ehle},
  {Erd}, {Gabriel}, {Guainazzi}, {Gondoin}, {Much}, {Munoz}, {Santos},
  {Schartel}, {Texier}, \& {Vacanti}}]{Jansen2001}
{Jansen}, F., et al. 2001, \aap, 365, L1

\bibitem[{{Kastner} {et~al.}(2002){Kastner}, {Huenemoerder}, {Schulz},
  {Canizares}, \& {Weintraub}}]{Kastner2002b}
{Kastner}, J.~H., {Huenemoerder}, D.~P., {Schulz}, N.~S., {Canizares}, C.~R.,
  \& {Weintraub}, D.~A. 2002, \apj, 567, 434

\bibitem[{{Kastner} {et~al.}(2004){Kastner}, {Richmond}, {Grosso}, {Weintraub},
  {Simon}, {Frank}, {Hamaguchi}, {Ozawa}, \& {Henden}}]{Kastner2004}
{Kastner}, J.~H., et al. 2004, \nat, 430, 429

\bibitem[{{Knude} \& {H$\phi$g}(1998)}]{Knude1998}
{Knude}, J., \& {H$\phi$g}, E. 1998, \aap, 338, 897

\bibitem[{{Koyama} {et~al.}(1996){Koyama}, {Hamaguchi}, {Ueno}, {Kobayashi}, \&
  {Feigelson}}]{Koyama1996}
{Koyama}, K., {Hamaguchi}, K., {Ueno}, S., {Kobayashi}, N., \& {Feigelson},
  E.~D. 1996, \pasj, 48, L87

\bibitem[{{Marti} {et~al.}(1995){Marti}, {Rodriguez}, \&
  {Reipurth}}]{Marti1995}
{Marti}, J., {Rodriguez}, L.~F., \& {Reipurth}, B. 1995, \apj, 449, 184

\bibitem[{{McClintock} \& {Remillard}(2003)}]{McClintock2003}
{McClintock}, J.~E., \& {Remillard}, R.~A. 2003, astro-ph/0306213

\bibitem[{{Mewe} {et~al.}(1995){Mewe}, {Kaastra}, \& {Liedahl}}]{Mewe1995}
{Mewe}, R., {Kaastra}, J.~S., \& {Liedahl}, D.~A. 1995, Legacy, 6, 16

\bibitem[{{Montmerle} {et~al.}(2000){Montmerle}, {Grosso}, {Tsuboi}, \&
  {Koyama}}]{Montmerle2000}
{Montmerle}, T., {Grosso}, N., {Tsuboi}, Y., \& {Koyama}, K. 2000, \apj, 532,
  1097

\bibitem[{{Morrison} \& {McCammon}(1983)}]{Morrison1983}
{Morrison}, R., \& {McCammon}, D. 1983, \apj, 270, 119

\bibitem[{{Pravdo} {et~al.}(2001){Pravdo}, {Feigelson}, {Garmire}, {Maeda},
  {Tsuboi}, \& {Bally}}]{Pravdo2001}
{Pravdo}, S.~H., {Feigelson}, E.~D., {Garmire}, G., {Maeda}, Y., {Tsuboi}, Y.,
  \& {Bally}, J. 2001, \nat, 413, 708

\bibitem[{{Rho} {et~al.}(2004){Rho}, {Ram\'{i}rez}, {Corcoran}, {Hamaguchi}, \&
  {Lefloch}}]{Rho2004}
{Rho}, J., {Ram\'{i}rez}, S.~V., {Corcoran}, M.~F., {Hamaguchi}, K., \&
  {Lefloch}, B. 2004, \apj, 607, 904

\bibitem[{{Saraceno} {et~al.}(1996){Saraceno}, {Andr\'{e}}, {Ceccarelli},
  {Griffin}, \& {Molinari}}]{Saraceno1996}
{Saraceno}, P., {Andr\'{e}}, P., {Ceccarelli}, C., {Griffin}, M., \&
  {Molinari}, S. 1996, \aap, 309, 827

\bibitem[{{Shibata} \& {Yokoyama}(2002)}]{Shibata2002}
{Shibata}, K., \& {Yokoyama}, T. 2002, \apj, 577, 422

\bibitem[{{Skinner} {et~al.}(2003){Skinner}, {Gagn{\' e}}, \&
  {Belzer}}]{Skinner2003}
{Skinner}, S., {Gagn{\' e}}, M., \& {Belzer}, E. 2003, \apj, 598, 375

\bibitem[{{Stelzer} {et~al.}(2000){Stelzer}, {Neuh{\" a}user}, \&
  {Hambaryan}}]{Stelzer2000}
{Stelzer}, B., {Neuh{\" a}user}, R., \& {Hambaryan}, V. 2000, \aap, 356, 949

\bibitem[{{Stern}(1998)}]{Stern1998}
{Stern}, R.~A. 1998, in ASP Conf. Ser. 154: Cool Stars, Stellar Systems, and
  the Sun, 223--+

\bibitem[{{Str{\" u}der} {et~al.}(2001){Str{\" u}der}, {Briel}, {Dennerl},
  {Hartmann}, {Kendziorra}, {Meidinger}, {Pfeffermann}, {Reppin}, {Aschenbach},
  {Bornemann}, {Br{\" a}uninger}, {Burkert}, {Elender}, {Freyberg}, {Haberl},
  {Hartner}, {Heuschmann}, {Hippmann}, {Kastelic}, {Kemmer}, {Kettenring},
  {Kink}, {Krause}, {M{\" u}ller}, {Oppitz}, {Pietsch}, {Popp}, {Predehl},
  {Read}, {Stephan}, {St{\" o}tter}, {Tr{\" u}mper}, {Holl}, {Kemmer},
  {Soltau}, {St{\" o}tter}, {Weber}, {Weichert}, {von Zanthier},
  {Carathanassis}, {Lutz}, {Richter}, {Solc}, {B{\" o}ttcher}, {Kuster},
  {Staubert}, {Abbey}, {Holland}, {Turner}, {Balasini}, {Bignami}, {La
  Palombara}, {Villa}, {Buttler}, {Gianini}, {Lain{\' e}}, {Lumb}, \&
  {Dhez}}]{Struder2001}
{Str{\" u}der}, L., et al. 2001, \aap, 365, L18

\bibitem[{{Townsley} {et~al.}(2000){Townsley}, {Broos}, {Garmire}, \&
  {Nousek}}]{Townsley2000ia}
{Townsley}, L.~K., {Broos}, P.~S., {Garmire}, G.~P., \& {Nousek}, J.~A. 2000,
  \apjl, 534, L139

\bibitem[{{Tsuboi} {et~al.}(2000){Tsuboi}, {Imanishi}, {Koyama}, {Grosso}, \&
  {Montmerle}}]{Tsuboi2000}
{Tsuboi}, Y., {Imanishi}, K., {Koyama}, K., {Grosso}, N., \& {Montmerle}, T.
  2000, \apj, 532, 1089

\bibitem[{{Tsuboi} {et~al.}(2001){Tsuboi}, {Koyama}, {Hamaguchi}, {Tatematsu},
  {Sekimoto}, {Bally}, \& {Reipurth}}]{Tsuboi2001}
{Tsuboi}, Y., {Koyama}, K., {Hamaguchi}, K., {Tatematsu}, K., {Sekimoto}, Y.,
  {Bally}, J., \& {Reipurth}, B. 2001, \apj, 554, 734

\bibitem[{{Tsuboi} {et~al.}(1998){Tsuboi}, {Koyama}, {Murakami}, {Hayashi},
  {Skinner}, \& {Ueno}}]{Tsuboi1998}
{Tsuboi}, Y., {Koyama}, K., {Murakami}, H., {Hayashi}, M., {Skinner}, S., \&
  {Ueno}, S. 1998, \apj, 503, 894

\bibitem[{{Tsujimoto} {et~al.}(2003){Tsujimoto}, {Koyama}, {Kobayashi}, {Goto},
  {Tsuboi}, \& {Tokunaga}}]{Tsujimoto2003}
{Tsujimoto}, M., {Koyama}, K., {Kobayashi}, N., {Goto}, M., {Tsuboi}, Y., \&
  {Tokunaga}, A.~T. 2003, \aj, 125, 1537

\bibitem[{{Tsujimoto} {et~al.}(2004){Tsujimoto}, {Koyama}, {Kobayashi},
  {Saito}, {Tsuboi}, \& {Chandler}}]{Tsujimoto2004}
{Tsujimoto}, M., {Koyama}, K., {Kobayashi}, N., {Saito}, M., {Tsuboi}, Y., \&
  {Chandler}, C.~J. 2004, \pasj, 56, 341

\bibitem[{{Turner} {et~al.}(2001){Turner}, {Abbey}, {Arnaud}, {Balasini},
  {Barbera}, {Belsole}, {Bennie}, {Bernard}, {Bignami}, {Boer}, {Briel},
  {Butler}, {Cara}, {Chabaud}, {Cole}, {Collura}, {Conte}, {Cros}, {Denby},
  {Dhez}, {Di Coco}, {Dowson}, {Ferrando}, {Ghizzardi}, {Gianotti}, {Goodall},
  {Gretton}, {Griffiths}, {Hainaut}, {Hochedez}, {Holland}, {Jourdain},
  {Kendziorra}, {Lagostina}, {Laine}, {La Palombara}, {Lortholary}, {Lumb},
  {Marty}, {Molendi}, {Pigot}, {Poindron}, {Pounds}, {Reeves}, {Reppin},
  {Rothenflug}, {Salvetat}, {Sauvageot}, {Schmitt}, {Sembay}, {Short},
  {Spragg}, {Stephen}, {Str{\" u}der}, {Tiengo}, {Trifoglio}, {Tr{\" u}mper},
  {Vercellone}, {Vigroux}, {Villa}, {Ward}, {Whitehead}, \&
  {Zonca}}]{Turner2001}
{Turner}, M.~J.~L., et al. 2001, \aap, 365, L27

\bibitem[{{Ueda} {et~al.}(1998){Ueda}, {Takahashi}, {Inoue}, {Tsuru}, {Sakano},
  {Ishisaki}, {Ogasaka}, {Makishima}, {Yamada}, \& {Ohta}}]{Ueda1998}
{Ueda}, Y., et al. 1998, \nat, 391, 866

\bibitem[{{van den Ancker}(1999)}]{Ancker1999b}
{van den Ancker}, M.~E. 1999, PhD thesis, Universiteit van Amsterdam

\bibitem[{{Weisskopf} {et~al.}(2002){Weisskopf}, {Brinkman}, {Canizares},
  {Garmire}, {Murray}, \& {Van Speybroeck}}]{Weisskopf2002}
{Weisskopf}, M.~C., {Brinkman}, B., {Canizares}, C., {Garmire}, G., {Murray},
  S., \& {Van Speybroeck}, L.~P. 2002, \pasp, 114, 1

\bibitem[{{Wilking} {et~al.}(1997){Wilking}, {McCaughrean}, {Burton}, {Giblin},
  {Rayner}, \& {Zinnecker}}]{Wilking1997}
{Wilking}, B.~A., {McCaughrean}, M.~J., {Burton}, M.~G., {Giblin}, T.,
  {Rayner}, J.~T., \& {Zinnecker}, H. 1997, \aj, 114, 2029, (W97)

\end{thebibliography}

\clearpage

\begin{deluxetable}{llrlccl}
\tablecolumns{7}
\tablewidth{0pc}
\tabletypesize{\scriptsize}
\tablecaption{Observation Logs\label{tbl:obslogs}}
\tablehead{
\colhead{Observation}&\colhead{Observatory}&\colhead{Seq. ID}&\colhead{Date}&\colhead{Exposure}&\colhead{$\Delta_{axis}$}&\colhead{Shift ($\Delta\alpha$,$\Delta\delta$)}\\
\colhead{}&\colhead{}&\colhead{}&&\colhead{(ksec)}&\colhead{}&\colhead{}
}
\startdata
\OBSP&\XMM&146390101&2003 Mar. 28&14.9/21.5&6\ARCMIN&(0\FARCS4, $-$1\FARCS0)\\
\OBSS&\XMM&146390201&2003 Mar. 29&18.1/24.0&6\ARCMIN&(0\FARCS2, $-$1\FARCS3)\\
\OBSCA&\CHANDRA&200017&2000 Oct. 7&19.7&2\ARCMIN&(0\FARCS03, 0\FARCS23)\\
\OBSCB&\CHANDRA&200194&2003 Jun. 26&37.6&0\FARCM2&(0\FARCS42, $-$0\FARCS03)\\
\enddata
\tablecomments{Exposure: EPIC pn/MOS for \XMM. $\Delta_{axis}$: Off-axis angle of the IRS7 region.
Objects used for the position correction: IRS2, IRS5, 
HBC677, CrA1, ISO-CrA136, HH101 IRS 4, ISO-CrA137, TY CrA 
and HD 176386. IRS1 and ISO-CrA134 were also used for the position 
correction of the \CHANDRA data.}
\end{deluxetable}

\begin{deluxetable}{clllccll}
\tablecolumns{8}
\tablewidth{0pc}
\tabletypesize{\tiny}
\tablecaption{Detected Sources\label{tbl:detsources}}
\tablehead{
\colhead{Source}&\colhead{Designation}&\colhead{Observation}&\colhead{($\alpha_{2000}$, $\delta_{2000}$)\tablenotemark{a}}&\colhead{Net cnts}&\colhead{$K$-band}&\colhead{Counterpart}\\
&&&\colhead{(h~m~s, d~\ARCMIN~\ARCSEC)}&\colhead{(cnts)}&\colhead{(mag)}
}
\startdata
\XB&XMMU J190156.3-365726&\OBSP&19~1~56.29, $-$36~57~26.5&467.7&$\gtrsim$19.4&10E\tablenotemark{d}, vdA~5(?)\tablenotemark{e}\\
   &			 &\OBSS&19~1~56.32, $-$36~57~26.4&1611.3&\nodata&\nodata\\
   &CXOU J190156.4-365728&\OBSCA&19~1~56.43, $-$36~57~27.7&19.6&\nodata&\nodata\\
   &			 &\OBSCB&19~1~56.39, $-$36~57~28.3&12.7&\nodata&\nodata\\
\XA&XMMU J190155.3-365721&\OBSPS&19~1~55.26, $-$36~57~20.8&114.0\tablenotemark{b}&12.2\tablenotemark{c}&10W\tablenotemark{d}, Source 4\tablenotemark{f}, IRS7, vdA~3(?)\tablenotemark{e}\\
   &CXOU J190155.3-365722&\OBSCA&19~1~55.34, $-$36~57~21.6&12.7&\nodata&\nodata\\
   &			 &\OBSCB&19~1~55.32, $-$36~57~21.8&22.4&\nodata&\nodata\\
\enddata
\tablecomments{Net cnts: EPIC pn + MOS for \XMM.}
\tablenotetext{a}{Positional uncertaintiess are $\sim$2\ARCSEC for \XMM and $\sim$1\ARCSEC for \CHANDRA.}
\tablenotetext{b}{Net counts in \OBSP.}
\tablenotetext{c}{Reference of the K$'$ magnitude of R1 (IRS~7) to \citet{Wilking1997}.}
\tablenotetext{d}{Reference to \citet{Brown1987} and \citet{Feigelson1998}.}
\tablenotetext{e}{Reference to \citet{Ancker1999b}.}
\tablenotetext{f}{Reference to \citet{Choi2004}.}
\end{deluxetable}

\begin{deluxetable}{lccccc}
\tablecolumns{6}
\tablewidth{0pc}
\tabletypesize{\scriptsize}
\tablecaption{Fitting Result of the \XB Light Curve \label{tbl:lcxmmxb}}
\tablehead{
\colhead{Observation}&\colhead{Binning}&\colhead{Model}&\colhead{Constant}&\colhead{Linear}&\colhead{$\chi^{2}$/d.o.f. (d.o.f.)}\\
&\colhead{(sec)}&&\colhead{(10$^{-2}$~\UNITCPS)}&\colhead{(10$^{-2}$~\UNITCPS day$^{-1}$)}}
\startdata
\OBSP&2,000&constant	&0.93&\nodata&1.63 (13)\\
\OBSS&1,000&constant	&5.3&\nodata&1.96 (26)\\
     &1,000&constant+linear	&3.7\tablenotemark{a}&9.3&1.21 (25)
\enddata
\tablenotetext{a}{Count rate at TJD = 12727.8564 day.}
\end{deluxetable}

\begin{deluxetable}{llclllllllc}
\tablecolumns{11}
\tablewidth{0pc}
\rotate
\tabletypesize{\tiny}
\tablecaption{Fitting Results of the Spectra\label{tbl:spec}}
\tablehead{
\colhead{Source}&\colhead{Observation}&\colhead{Model}&\colhead{Comp.}&\colhead{\NH}&\colhead{\KT}&\colhead{Abundance}&\colhead{log \EM}&\colhead{Flux$_{6.4 keV}$}&\colhead{$\chi^{2}$/d.o.f. (d.o.f)}&\colhead{log \LX}\\
&&&&\colhead{(10$^{22}$\UNITNH)}&\colhead{(keV)}&\colhead{(solar)}&\colhead{(\UNITEI)}&\colhead{(10$^{-6}$ \UNITPFLUX)}&&\colhead{(\UNITLUMI)}}
\startdata
\XB&\OBSP&A(1T)&&13.2 (9.6$-$18.3)&5.1 (2.8$-$10.4)&0.5 (0.2$-$1.4) &53.2 (53.0$-$53.6)&2.4&1.22 (21)&30.4\\
      &\OBSS&B(1T)&&24.7 (21.6$-$28.3)&4.4 (3.4$-$6.1)&0.2 (0.1$-$0.3) &54.1 (54.0$-$54.1)&3.4&0.99 (66)&31.2\\
      &\OBSPS&C(2T)&Var$_{\rm 1}$&28.1 (22.5$-$33.8)&2.7 (1.8$-$4.2)&0.2 (0.1$-$0.3) &53.8 (53.5$-$54.1)&2.8 (0.8$-$4.8)&0.97 (86)&30.8\\
      & &	     &Var$_{\rm 2}$&=Var$_{\rm 1}$&4.0 (2.9$-$5.3)&=Var$_{\rm 1}$&54.1 (54.0$-$54.4)&3.4 (1.4$-$5.5)&\nodata&31.2\\
      & &     &Const.&4.2 (1.4$-$11.4)&2.3 (0.4$-$)&=Var$_{\rm 1}$&52.5 (51.8$-$53.5)&\nodata&\nodata&29.6\\
\XA&\OBSP&D(1T)&&33.7 (19.1$-$60.0)&4.7 (1.6$-$)&0.3 (fix) &53.3 (52.8$-$54.3)&\nodata&0.65 (8)&30.5\\
&\OBSCAB&E(1T)&&33.7 (fix) & 4.7 (fix) & 0.3 (fix)&52.9 (52.8$-$53.0)&\nodata&0.60 (3)&30.0\\
\enddata
\tablecomments{Var$_{1(2)}$: variable hard component in \OBSP (\OBSS).
Const.: constant soft component.
\LX: absorption corrected X-ray luminosity in the 0.5$-$10 keV band. Distance assumes $d \sim$170 pc.}
\end{deluxetable}

\clearpage

\clearpage

\clearpage

\clearpage

\clearpage

\clearpage

\clearpage

\clearpage

\end{document}